\documentclass[10pt,twocolumn,letterpaper]{article}
\usepackage{cvpr}
\usepackage[breaklinks,colorlinks]{hyperref}

\title{Interactive Multi-Turn Retrieval for Health Videos}

\author{\begin{tabular}{c}
Chengzheng Wu$^{1}$ \quad Ke Qiu$^{2}$ \quad Baoming Zhang$^{3}$ \quad Ruiyu Mao$^{4}$\\
Xulong Tang$^{2}$ \quad Kaixing Yang$^{5}$\\[0.35em]
$^{1}$Wuxi Dipont School of Arts and Science \quad
$^{2}$Malou Tech \quad
$^{3}$Chongqing University\\
$^{4}$Case Western Reserve University \quad
$^{5}$Renmin University of China\\[0.25em]
{\tt\small 10felix.wu01@gmail.com \quad ke.qiu@maloutech.com \quad baomingzhang@alu.cqu.edu.cn}\\
{\tt\small rxm603@case.edu \quad xulong.tang@maloutech.com \quad yangkaixing@ruc.edu.cn}
\end{tabular}}

\begin{document}
\maketitle

\begin{abstract}
The growing availability of health-related instructional videos creates new opportunities for clinical training, patient rehabilitation, and health education, yet existing retrieval systems remain largely single-turn: a user submits one query and receives one ranked list. This interaction is brittle in health scenarios, where information needs are often vague at first and become clinically meaningful only after follow-up constraints such as posture, hand placement, contraindications, equipment, or patient condition are specified. We introduce interactive multi-turn semantic retrieval for health videos and construct MHVRC, a Multi-Turn Health Video Retrieval Corpus, by combining video-grounded descriptions from VideoChat-Flash with query refinements generated by DeepSeek. We further propose DATR, a Dialogue-Aware Two-Stage Retrieval framework. DATR first performs efficient coarse retrieval with a CLIP-style dual encoder and sparse frame sampling, then re-ranks the top candidates through multi-turn query fusion and a lightweight cross-encoder scoring module. Experiments on MHVRC show consistent gains over strong text-video retrieval baselines, while user studies indicate that refined multi-turn queries better capture fine-grained procedural semantics than single-turn annotations. The work establishes a benchmark and a scalable technical recipe for interactive health video retrieval.
\end{abstract}

\section{Introduction}

\begin{figure*}[t]
  \centering
  \includegraphics[width=0.82\textwidth]{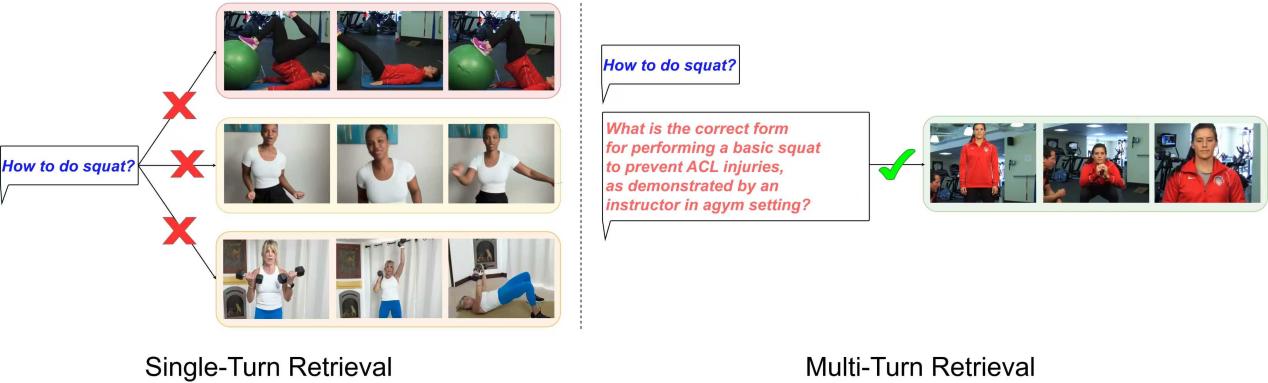}
  \caption{Single-turn versus multi-turn health video retrieval. Single-turn search often matches broad topic words, while interactive refinement can expose procedural details such as body posture, safety constraints, and target clinical context.}
  \label{fig:single_multi}
\end{figure*}

Video retrieval has advanced rapidly with vision-language pretraining, contrastive objectives, and scalable transformer encoders \citep{radford2021clip,luo2021clip4clip,bain2021frozen,lei2021clipbert}. These methods make it possible to align natural language with visual content across large collections, and they have been validated on general-domain benchmarks such as MSR-VTT, YouCook2, ActivityNet Captions, DiDeMo, and TVR \citep{xu2016msrvtt,zhou2018youcook2,krishna2017activitynet,hendricks2017didemo,lei2020tvr}. However, health video retrieval is more demanding than generic text-video matching. Users often seek procedural, safety-sensitive, and personally situated information: a clinician may look for an exercise demonstration appropriate for a specific injury; a patient may ask how to perform a maneuver safely at home; an educator may need a clear procedural segment for training. In such contexts, a single query is rarely sufficient.

Figure~\ref{fig:single_multi} illustrates the central gap. A query such as ``how to do a squat'' may retrieve generally relevant exercise videos, yet a useful clinical answer may require follow-up constraints: knee alignment, anterior cruciate ligament injury prevention, instructor demonstration, or appropriate difficulty. The retrieval system must therefore model an evolving information need rather than a static sentence. This motivates the task of interactive multi-turn health video retrieval, where a user provides an initial query $q_1$ and one or more refinements $q_t$, and the model must return videos that satisfy the latest intent while preserving the context established earlier.

Existing work provides useful building blocks but does not solve this setting. General text-video retrieval models excel at coarse alignment but usually ignore conversational history \citep{luo2021clip4clip,bain2021frozen,lei2021clipbert,yu2020hero}. Conversational information retrieval studies query reformulation and context tracking, yet these systems are rarely grounded in long instructional videos or health-specific procedural semantics \citep{yu2021cir,nogueira2019bert}. Medical video datasets such as MedVidQA enable health video question answering, but their questions are single-turn and often too coarse to represent real iterative search \citep{abacha2022medvidqa}. Health AI work in rehabilitation and surgical video understanding shows the practical importance of vision-based procedural analysis \citep{bae2019rehab,li2020smartrehab,twinanda2017endonet,jin2018svrcnet}, but it has not produced a benchmark for dialogue-aware retrieval.

To address these limitations, we introduce MHVRC, a Multi-Turn Health Video Retrieval Corpus. MHVRC is derived from trusted instructional health videos and pairs each video with an initial coarse query, a detailed video-grounded description, and a refined follow-up query. VideoChat-Flash is used to generate procedural descriptions that emphasize visual evidence such as posture, body contact, equipment, and action sequence; DeepSeek is then prompted to reformulate the initial query into a realistic follow-up conditioned on the video description. This pipeline turns coarse single-turn supervision into multi-turn retrieval data while retaining procedural grounding.

We further propose DATR, a Dialogue-Aware Two-Stage Retrieval framework. Stage I uses a dual encoder to map text and videos into a shared embedding space, enabling scalable coarse retrieval over the full database. Stage II fuses the initial and refined queries, applies a lightweight cross-encoder over the Stage-I top-$K$ candidates, and re-ranks them according to fine-grained procedural relevance. DATR balances efficiency and precision: the dual encoder avoids exhaustive cross-encoding over the full corpus, while the re-ranker captures intent shifts that single-turn retrieval misses.

Our contributions are:
\begin{itemize}
  \item We formulate interactive multi-turn health video retrieval as a distinct task that models progressive refinement of user intent in clinically meaningful search scenarios.
  \item We construct MHVRC, a multi-turn corpus built from health instructional videos, video-grounded descriptions, and refined follow-up queries.
  \item We propose DATR, a dialogue-aware two-stage retrieval framework combining efficient dual-encoder retrieval with query-fusion re-ranking.
  \item We provide quantitative, ablation, qualitative, and user-study evidence showing that multi-turn modeling improves retrieval quality and better reflects practical health information seeking.
\end{itemize}

\section{Related Work}

\subsection{Human-Centered AI}
Human-centered AI increasingly treats the body as a carrier of action, identity, intention, and social context rather than as an isolated visual object. In video understanding, large-scale vision-language models, temporal transformers, and skeleton-based action models provide complementary representations for actions, events, instructions, and moment grounding \citep{radford2021clip,arnab2021vivit,bertasius2021timesformer,zhang2025robust2d,lei2021clipbert,krishna2017activitynet,lei2020tvr}. These representations support applications such as search, education, assistive interfaces, and embodied interaction, where a system must connect visible human behavior with language-level intent.

Digital-human generation follows two broad but related lines. One line focuses on human video generation and avatar animation, where pose transfer, identity consistency, audio-visual synchronization, and controllable appearance are central challenges \citep{chang2023magic,hu2023animate,yang2025macedance,uniavgen2025,zhang2024dancecamera3d}. Another line focuses on 3D motion generation, where methods generate or edit skeleton-level motion from text, speech, music, or other control signals \citep{guo2022tm2t,tevet2023mdm,chen2023executing,jiang2023motiongpt,tang2023motionbert}. Within 3D motion, text-to-motion emphasizes semantic controllability, speech-to-gesture emphasizes communicative timing and emphasis, and audio-conditioned motion uses rhythm or prosody to preserve temporal structure \citep{zhang2023tm2d,zhang2025semtalk,zhang2025echomask,li2021ai,siyao2022bailando}. Recent dance and conducting systems can be viewed as representative stress tests for these ideas because they require style control, beat alignment, and fine-grained temporal coordination \citep{yang2024beatdance,yang2024megadance,yang2024flowerdance,zhang2024bitdiff,personaldance2024}. Although our task is retrieval rather than generation, it shares the same human-centered premise: systems should respond to bodily detail, user capability, and progressively specified intent.

\subsection{Retrieval}
Text-video retrieval aims to rank videos according to natural-language queries. Early benchmark-driven progress relied on paired video-caption datasets that exposed models to diverse actions, scenes, and temporal descriptions \citep{xu2016msrvtt,zhou2018youcook2,krishna2017activitynet,hendricks2017didemo,lei2020tvr}. CLIP-style pretraining then made large-scale contrastive alignment a dominant recipe for efficient retrieval \citep{radford2021clip,luo2021clip4clip,bain2021frozen,lei2021clipbert}. Transformer-based video encoders improved temporal modeling and sparse frame reasoning, making it practical to encode longer videos without dense processing of every frame \citep{arnab2021vivit,bertasius2021timesformer,yu2020hero,lei2021clipbert}. Hierarchical and multimodal pretraining methods further combine local frame evidence, global video context, and textual supervision for stronger first-stage retrieval \citep{yu2020hero,lei2020tvr,lei2018tvqa,bain2021frozen}.

Retrieval has also expanded from generic caption matching to more structured forms of alignment. Moment retrieval and video QA require models to connect language with temporal segments rather than whole clips \citep{hendricks2017didemo,krishna2017activitynet,lei2020tvr,lei2018tvqa}. Domain-sensitive retrieval shows that global semantics may be insufficient when local temporal cues carry the actual intent, as seen in procedure learning and music-dance matching \citep{zhou2018youcook2,yang2024beatdance,zhang2023tm2d,li2021ai}. Re-ranking methods complement dual encoders by applying richer query-candidate interaction after efficient coarse search \citep{nogueira2019bert,luo2021clip4clip,yu2020hero}. Conversational search further argues that users often refine underspecified needs through interaction, making query history, reformulation, and context tracking important for ranking \citep{yu2021cir,nogueira2019bert,lei2018tvqa}. Recent multi-turn and dialogue-aware retrieval work therefore motivates a shift from one-shot matching to intent evolution, where the current query must be interpreted together with previous turns \citep{yu2021cir,lei2020tvr,lei2018tvqa,demner2009clinical}. Health video retrieval pushes this direction further because refinements may involve posture, contraindications, equipment, or rehabilitation stage rather than generic visual attributes \citep{abacha2022medvidqa,bae2019rehab,li2020smartrehab,wang2023videorehab}.

\subsection{Health AI}
Health-related AI has moved from diagnostic classification toward systems that support education, rehabilitation, therapy, and patient self-management. Medical video QA and misinformation retrieval demonstrate the need for trustworthy multimedia access, while clinical NLP and decision-support research show that health information needs are often contextual and risk-sensitive \citep{abacha2022medvidqa,demner2009clinical,li2020smartrehab,wang2023videorehab}. In rehabilitation, computer vision has been used to assess exercise quality, detect pose errors, and provide feedback for remote or home-based training \citep{bae2019rehab,li2020smartrehab,wang2023videorehab,tang2023motionbert}. In surgical and procedural settings, models such as EndoNet and SV-RCNet recognize operative phases and workflow states, highlighting the importance of temporal structure and domain-specific action semantics \citep{twinanda2017endonet,jin2018svrcnet,arnab2021vivit,bertasius2021timesformer}.

Despite this progress, health video access remains underexplored as an interactive retrieval problem. A patient may need to refine an exercise by pain level, body side, equipment, age, or contraindication; a therapist may need a demonstration that matches a rehabilitation phase; an educator may search for a particular procedural step rather than an entire video. Single-turn retrieval can rank broadly relevant clips, but it does not model the collaborative process by which users clarify what is safe and useful. MHVRC and DATR are designed for this setting. They connect video-language retrieval with human-centered health AI by making iterative query refinement a first-class part of the benchmark and model.

\section{MHVRC Dataset}

\begin{figure*}[t]
  \centering
  \includegraphics[width=0.90\textwidth]{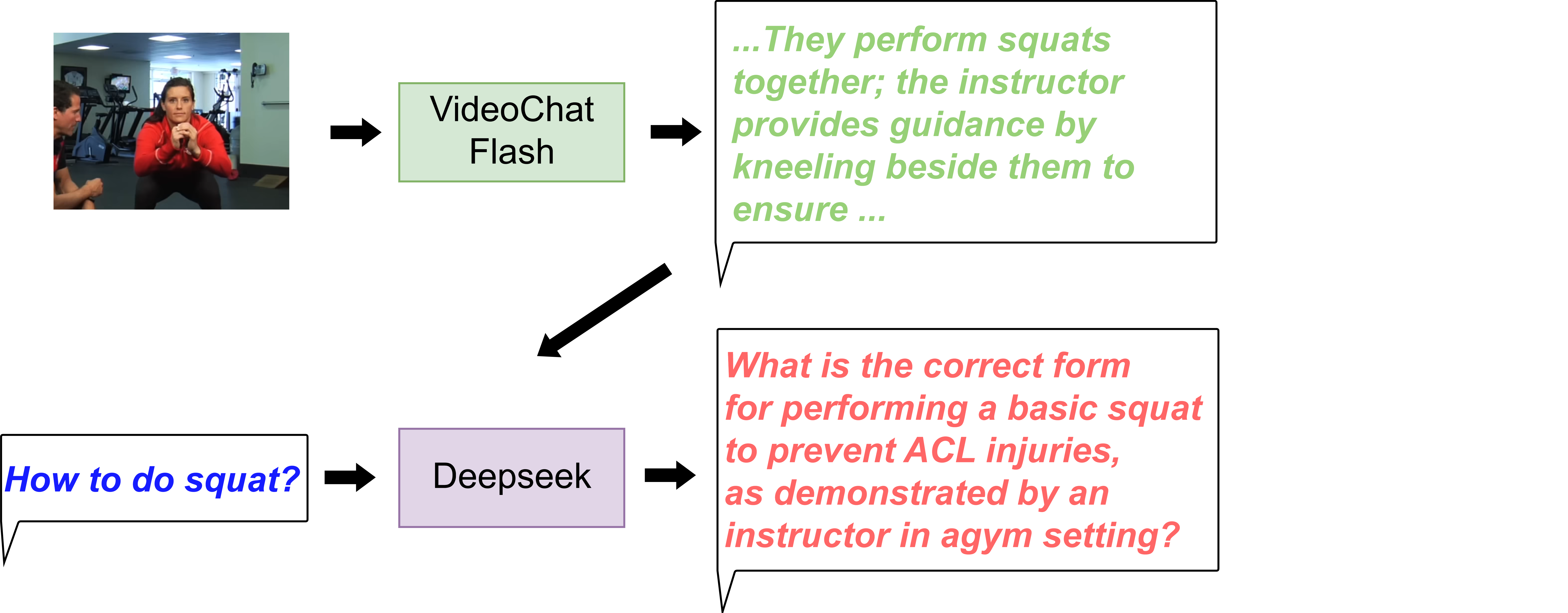}
  \caption{Construction pipeline of MHVRC. VideoChat-Flash produces procedural descriptions from health videos, while DeepSeek refines coarse user-style queries into specific follow-up queries grounded in the generated descriptions.}
  \label{fig:dataset_pipeline}
\end{figure*}

\subsection{Motivation and Source Limitation}
MedVidQA is an important medical-video benchmark with manually curated questions and timestamped answers from authoritative medical videos \citep{abacha2022medvidqa}. However, its annotations reflect a single-turn question-answering setting. Queries such as ``how to perform Epley maneuver for vertigo'' identify a topic, but they do not express the detailed needs that arise in real search, such as patient orientation, head rotation, hand placement, or the moment at which the patient should be moved. For rehabilitation and clinical training, these details are not cosmetic; they determine whether a retrieved video is practically useful.

MHVRC is designed to preserve the credibility of health instructional sources while adding multi-turn interaction. Each data item is represented as a triplet:
\begin{equation}
  (q_1, d_v, q_2),
  \label{eq:triplet}
\end{equation}
where $q_1$ is an initial coarse query, $d_v$ is a video-grounded procedural description, and $q_2$ is a refined follow-up query. The construction process is summarized in Fig.~\ref{fig:dataset_pipeline}.

\subsection{Video-Grounded Description}
For each video, we use VideoChat-Flash to generate a detailed procedural description. The prompt instructs the model to behave as a medical video analyst and to describe body posture, hand placement, medical actions, equipment, patient-instructor interaction, and step order. We use deterministic decoding to reduce variance and allocate a long frame budget so that the model can observe multiple procedural phases. For an Epley maneuver video, for example, the generated description identifies the supine patient, the instructor's two-hand support around the head and neck, the rotation sequence, and the stabilizing hand positions. Such descriptions provide a richer semantic bridge than short captions.

\subsection{Multi-Turn Query Refinement}
DeepSeek is then prompted to reformulate the initial query conditioned on the video description. The prompt emphasizes specificity, clinical relevance, and evidence grounded in the observed video. A coarse query such as ``how to perform Epley maneuver for vertigo'' may be refined to ``how is the Epley maneuver performed on a supine patient, with specific hand placement on the head and neck?'' This refinement simulates the second turn of a realistic search session, where a user narrows the intent after seeing or describing an initial result.

\subsection{Statistics and Advantages}
MHVRC contains approximately 3,000 query triplets derived from about 900 instructional medical videos, with more than 90 hours of video content. The videos cover first aid and emergency response, musculoskeletal rehabilitation, cardiovascular monitoring, and patient self-management. The refined queries are longer and more specific than the initial queries, typically adding procedural constraints, visual details, or safety-relevant context.

Compared with single-turn annotations, MHVRC has four advantages. First, it captures fine-grained procedural semantics such as posture and hand placement. Second, it explicitly represents query evolution. Third, the refinement process produces more realistic health information needs than topic-only questions. Fourth, the generated descriptions connect the query text to visual evidence, which supports multimodal representation learning and retrieval evaluation.

\subsection{Annotation Quality Control}
Although MHVRC is built with automatic vision-language and language-model components, its annotation design is intended to reduce the most common failure modes of synthetic data. First, the video description step is constrained to observable evidence. Prompts ask the model to describe visible actions, body configuration, equipment, and spatial relations rather than infer diagnosis or give medical recommendations. This distinction is important because a retrieval benchmark should encode what appears in the video, not unsupported clinical advice. Second, query refinement is conditioned on both the initial query and the grounded description. The follow-up query therefore tends to add concrete constraints, such as ``supine patient,'' ``supporting the neck,'' or ``knee alignment,'' rather than drifting to unrelated topics.

We also organize the dataset around procedural granularity. A useful health query often has three semantic layers: the broad medical or educational topic, the visible action sequence, and the user-specific constraint. For example, ``Epley maneuver'' is a topic, ``rotate the patient's head while supine'' is a visible procedure, and ``specific hand placement on the head and neck'' is a retrieval constraint. MHVRC explicitly represents this hierarchy through $(q_1,d_v,q_2)$. This structure makes the dataset suitable for evaluating whether a model can retrieve beyond topic similarity.

Finally, the dataset is intended to be auditable. Each refined query can be traced back to a generated description, and each description can be checked against the corresponding video. This traceability is valuable for health applications because a failure can be categorized: the description may omit an important visual detail, the query may over-specify the procedure, or the retriever may fail despite correct annotations. Such decomposition makes MHVRC more useful than a flat set of query-video pairs when diagnosing model behavior.

\section{Method}

\begin{figure}[t]
  \centering
  \includegraphics[width=0.98\linewidth]{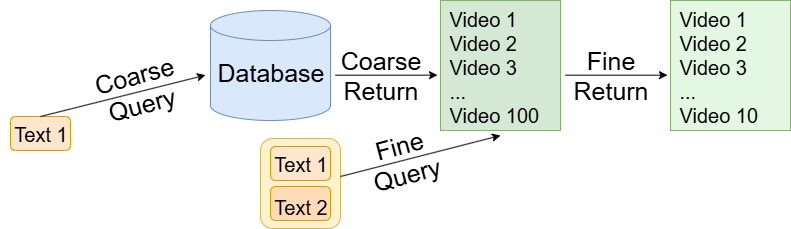}
  \caption{Overview of DATR. Stage I retrieves candidates with a scalable dual encoder; Stage II fuses multi-turn queries and re-ranks the top candidates.}
  \label{fig:framework}
\end{figure}

\subsection{Problem Definition}
Let the video collection be
\begin{equation}
  \mathcal{V} = \{v_i\}_{i=1}^{N},
  \label{eq:collection}
\end{equation}
where each $v_i$ is a health instructional video. A user provides a query sequence
\begin{equation}
  Q_t = \{q_1, q_2, \ldots, q_t\},
  \label{eq:query_sequence}
\end{equation}
where $q_1$ is the initial query and $q_t$ denotes the latest refinement. The goal is to return a ranked list
\begin{equation}
  \mathcal{R}_t = \operatorname{rank}(\mathcal{V}\mid Q_t)
  \label{eq:rank}
\end{equation}
that satisfies the refined intent at turn $t$. The model must solve three coupled challenges: semantic granularity, intent evolution, and scalability.

\subsection{Framework Overview}
DATR decomposes retrieval into coarse search and fine re-ranking. Stage I computes
\begin{equation}
  \mathcal{R}_1 = \operatorname{TopK}_{v\in\mathcal{V}}
  \operatorname{sim}\!\left(f_T(q_1), f_V(v)\right),
  \label{eq:stage1}
\end{equation}
where $f_T$ and $f_V$ are text and video encoders and $\operatorname{sim}$ is cosine similarity. Stage II fuses the multi-turn queries and re-ranks only the Stage-I candidates:
\begin{equation}
  \mathcal{R}_2 =
  \operatorname{TopM}_{v\in\mathcal{R}_1}
  \operatorname{score}\!\left(f_F(q_1,q_2), f_V(v)\right).
  \label{eq:stage2}
\end{equation}
Here $f_F$ is a fusion encoder that captures both continuity with $q_1$ and new constraints introduced by $q_2$.

\begin{figure*}[t]
  \centering
  \includegraphics[width=0.82\textwidth]{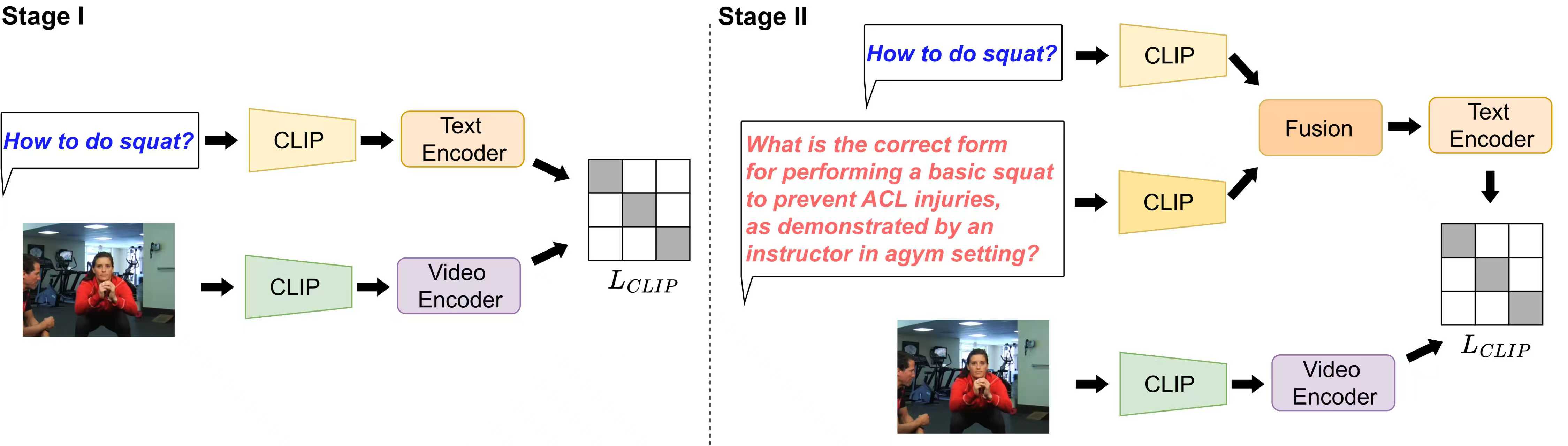}
  \caption{Architecture of the two-stage retrieval process. The wide pipeline is placed as a double-column figure to preserve readability.}
  \label{fig:architecture}
\end{figure*}

\subsection{Stage I: Coarse Retrieval}
The text encoder processes $q_1$ with a CLIP-style transformer and maps it to a shared embedding:
\begin{equation}
  z_T = f_T(q_1)\in\mathbb{R}^{d}.
  \label{eq:text_embed}
\end{equation}
To adapt the representation to health-domain retrieval while keeping the model lightweight, we apply a feed-forward adapter:
\begin{equation}
  \tilde{z}_T = W_2\,\sigma(W_1 z_T + b_1) + b_2,
  \label{eq:adapter}
\end{equation}
where $\sigma(\cdot)$ denotes ReLU activation.

For each video, we uniformly sample 32 frames and encode them as
\begin{equation}
  X = [x_1,\ldots,x_{32}]\in\mathbb{R}^{32\times d}.
  \label{eq:frames}
\end{equation}
The sampled frame embeddings are processed by $L=6$ transformer layers. For hidden state $H$,
\begin{equation}
  \operatorname{Attn}(H)=
  \operatorname{softmax}\!\left(\frac{QK^\top}{\sqrt{d}}\right)V,
  \label{eq:attention}
\end{equation}
followed by the feed-forward block
\begin{equation}
  \operatorname{FFN}(H)=W_2\sigma(W_1H+b_1)+b_2.
  \label{eq:ffn}
\end{equation}
To reduce redundant temporal tokens, a 1D convolutional downsampler is applied after each transformer block:
\begin{equation}
  H^{(\ell+1)} =
  \operatorname{Conv1D}_{k=3,s=2}\!\left(\operatorname{Transformer}(H^{(\ell)})\right).
  \label{eq:downsample}
\end{equation}
The final video representation is mean pooled:
\begin{equation}
  z_V = \operatorname{Pool}(H^{(L)}).
  \label{eq:video_embed}
\end{equation}

Stage I is trained with a bidirectional CLIP-style contrastive loss:
\begin{align}
  \mathcal{L}_{t2v} &=
  -\frac{1}{B}\sum_{i=1}^{B}
  \log
  \frac{\exp(\operatorname{sim}(z^i_T,z^i_V)/\tau)}
  {\sum_{j=1}^{B}\exp(\operatorname{sim}(z^i_T,z^j_V)/\tau)}, \\
  \mathcal{L}_{v2t} &=
  -\frac{1}{B}\sum_{i=1}^{B}
  \log
  \frac{\exp(\operatorname{sim}(z^i_V,z^i_T)/\tau)}
  {\sum_{j=1}^{B}\exp(\operatorname{sim}(z^i_V,z^j_T)/\tau)}, \\
  \mathcal{L}_{\mathrm{clip}} &= \frac{1}{2}(\mathcal{L}_{t2v}+\mathcal{L}_{v2t}),
  \label{eq:clip_loss}
\end{align}
where $B$ is the batch size and $\tau$ is a learnable temperature.

\subsection{Stage II: Fine Re-Ranking}
Stage II models the refined intent by combining the embeddings of $q_1$ and $q_2$:
\begin{equation}
  z_F =
  \operatorname{MLP}\left([z_{q_1};z_{q_2};z_{q_1}+z_{q_2};z_{q_1}\odot z_{q_2}]\right),
  \label{eq:fusion}
\end{equation}
where $[\cdot;\cdot]$ denotes concatenation and $\odot$ is element-wise multiplication. The additive term preserves topic continuity, while the multiplicative term highlights new constraints.

Each candidate $v\in\mathcal{R}_1$ is scored as
\begin{equation}
  s(v,Q_2)=w^\top \phi([z_F;z_V;z_F\odot z_V]) + b,
  \label{eq:rerank_score}
\end{equation}
where $\phi$ is a lightweight cross-encoder module. The final output is
\begin{equation}
  \mathcal{R}_{\mathrm{final}} =
  \operatorname{TopM}_{v\in\mathcal{R}_1}s(v,Q_2).
  \label{eq:final_rank}
\end{equation}
By restricting cross-encoding to the top 100 candidates, DATR captures procedural nuances while avoiding the cost of full-database re-ranking.

\subsection{Implementation Details}
Unless otherwise specified, all embeddings use dimension $d=512$. The video branch samples 32 frames uniformly from each instructional video. Uniform sampling is a conservative choice for this first benchmark: it avoids requiring shot boundaries or action proposals and makes the evaluation reproducible across videos of different lengths. Each sampled frame is encoded by a visual backbone initialized from vision-language pretraining, then passed through the temporal transformer described above. The transformer uses $L=6$ layers and $h=8$ attention heads. Temporal downsampling is applied after each block so that the model can capture long-range context without preserving all frame tokens until the final layer.

The text branch encodes the initial and refined queries independently before fusion. We use the same tokenizer and base transformer for $q_1$ and $q_2$, which avoids introducing turn-specific parameters and encourages both queries to live in the same semantic space. The adapter in Eq.~\eqref{eq:adapter} is lightweight and can be viewed as a domain calibration module. It lets the model learn health-specific alignment, such as the difference between a general exercise query and a rehabilitation-oriented query, without discarding the broad visual-language knowledge learned during pretraining.

For Stage I, all video embeddings can be precomputed and indexed. Retrieval is therefore efficient at inference time: a query embedding is compared against the database using cosine similarity, and the top-$K$ candidates are passed to Stage II. In our experiments we use $K=100$ because it maintains high candidate recall while keeping the cross-encoder cost manageable. Stage II is trained with the same positive video as Stage I but sees harder negatives sampled from the candidate pool. This training design encourages the re-ranker to focus on fine distinctions such as posture, equipment, and procedural phase rather than broad topic matching.

\subsection{Why Two Stages Are Necessary}
A single dual encoder is attractive because it is scalable, but it compresses the entire query and video into one vector each. This is often sufficient for general topical retrieval, yet health videos require more local distinctions. Two videos may both demonstrate squats, but only one may emphasize injury prevention, a particular knee angle, or a beginner-safe form. These details can be diluted in a global embedding. Conversely, a full cross-encoder over the entire database would be too expensive for interactive use, especially when users refine queries repeatedly.

DATR separates these responsibilities. Stage I is optimized for coverage: it should include relevant videos in the candidate set even if the ranking is imperfect. Stage II is optimized for precision: it can examine richer query-video interactions over a much smaller set. The fusion representation in Eq.~\eqref{eq:fusion} is particularly important because a follow-up query does not replace the initial query. Instead, it modifies it. Additive fusion preserves continuity with the original intent, while multiplicative fusion highlights constraints that distinguish the current turn from the previous one. This design is simple, but the ablations show that both terms contribute to retrieval quality.

\section{Experiments}

\subsection{Evaluation Protocol}
We evaluate on MHVRC using standard text-to-video retrieval metrics: Recall@$K$ for $K\in\{1,5,10,50,100\}$, Median Rank (MedR), and Mean Rank (MeanR). Recall@$K$ measures whether the ground-truth video appears in the top $K$ results, while MedR and MeanR summarize the rank distribution. We split videos into training and testing subsets with an 8:2 ratio and ensure that videos from the same source do not cross the split.

We report both strict and relaxed metrics because they reflect different use cases. R@1 and R@5 matter when the system is used as a direct assistant and users expect the first few results to be usable. R@50 and R@100 matter for two-stage systems because they indicate whether a first-stage retriever provides enough coverage for later re-ranking. MedR and MeanR complement recall by revealing whether failures are near misses or complete ranking breakdowns. In health video retrieval, near misses can still be useful for browsing, but complete failures are more problematic because they may send the user to unrelated or unsafe demonstrations.

\subsection{Baselines}
We choose baselines that cover the main design families in text-video retrieval. CLIP4Clip represents direct adaptation of CLIP-style contrastive alignment to video. Frozen-in-Time jointly models images and videos, providing a strong general-purpose retrieval backbone. CLIPBERT emphasizes sparse frame sampling and efficient video-language learning, making it relevant to long instructional videos. HERO uses hierarchical pretraining to integrate local and global video-language information. All baselines are evaluated in the same single-turn setting using the available query text; DATR differs by explicitly exploiting both $q_1$ and $q_2$.

This comparison is intentionally conservative. We do not claim that the baselines are weak models; rather, they are strong retrieval systems designed for a different interaction assumption. Their failure mode on MHVRC is not a lack of visual-language capacity in general, but a lack of dialogue-aware intent modeling. This distinction matters because it suggests that future work can combine stronger pretrained backbones with the multi-turn structure introduced here.

\subsection{Comparison with Baselines}
\begin{table}[t]
  \centering
  \caption{Text-to-video retrieval results on MHVRC. Higher Recall and lower rank are better.}
  \label{tab:main}
  \resizebox{\linewidth}{!}{
  \begin{tabular}{lrrrrrrr}
    \toprule
    Model & R@1 & R@5 & R@10 & R@50 & R@100 & MedR & MeanR\\
    \midrule
    CLIP4Clip & 13.8 & 34.2 & 46.3 & 77.9 & 86.0 & 13 & 37.2\\
    Frozen-in-Time & 12.6 & 32.5 & 44.1 & 75.6 & 84.2 & 14 & 39.8\\
    CLIPBERT & 14.7 & 36.1 & 48.5 & 78.3 & 87.1 & 12 & 35.0\\
    HERO & 15.2 & 37.5 & 49.8 & 80.0 & 88.4 & 11 & 32.7\\
    DATR (ours) & \textbf{19.5} & \textbf{44.8} & \textbf{57.1} & \textbf{85.6} & \textbf{92.7} & \textbf{7} & \textbf{25.6}\\
    \bottomrule
  \end{tabular}}
\end{table}

Table~\ref{tab:main} compares DATR with CLIP4Clip, Frozen-in-Time, CLIPBERT, and HERO. These baselines represent strong single-turn retrieval paradigms: CLIP4Clip transfers CLIP alignment to video, Frozen-in-Time jointly trains image and video encoders, CLIPBERT uses sparse sampling for efficient video-language learning, and HERO uses hierarchical video-language pretraining. DATR outperforms all baselines across Recall and rank metrics. The improvement is largest at stricter cutoffs, with R@1 increasing from 15.2 for HERO to 19.5 and R@10 increasing from 49.8 to 57.1. This suggests that multi-turn fusion is particularly helpful when the system must identify the most relevant procedural video rather than merely include it somewhere in a broad candidate set.

\subsection{Qualitative Analysis}
\begin{figure*}[t]
  \centering
  \includegraphics[width=0.78\textwidth]{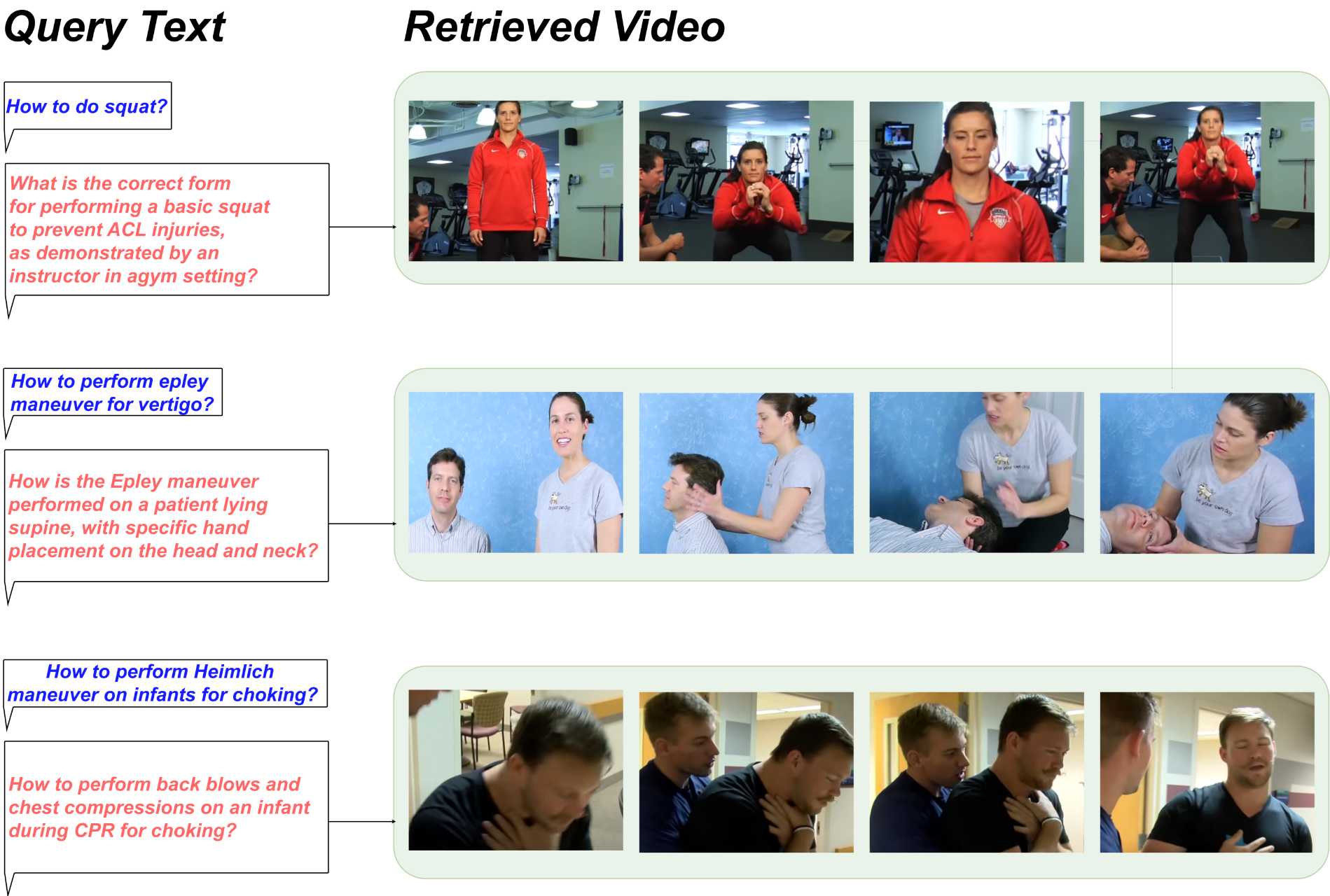}
  \caption{Qualitative retrieval examples from DATR. Retrieved videos match refined health queries involving exercise, rehabilitation, and procedural demonstrations.}
  \label{fig:qualitative}
\end{figure*}

Figure~\ref{fig:qualitative} shows representative retrieval outputs. The examples demonstrate that DATR can handle diverse query types, from general exercise instruction to more specific rehabilitation procedures. When refined queries include patient posture, injury prevention, or hand-placement details, DATR returns videos aligned with the intended procedural context. The qualitative results complement the quantitative gains: the model is not only improving aggregate rank metrics, but also surfacing examples that appear clinically meaningful and useful for instruction.

\subsection{User Study}
We conducted two user studies. The first evaluates query refinement quality. We sampled 100 query pairs and asked ten participants with healthcare or HCI background to rate the initial query and refined query on specificity, clarity, clinical relevance, and retrieval utility using a 5-point Likert scale.

\begin{table}[t]
  \centering
  \caption{User evaluation of query refinement.}
  \label{tab:query_user}
  \resizebox{\linewidth}{!}{
  \begin{tabular}{lrrr}
    \toprule
    Dimension & $q_1$ & $q_2$ & Improvement\\
    \midrule
    Specificity & 2.4 & 4.5 & +2.1\\
    Clarity & 3.0 & 4.6 & +1.6\\
    Clinical relevance & 2.8 & 4.7 & +1.9\\
    Retrieval utility & 2.9 & 4.8 & +1.9\\
    \bottomrule
  \end{tabular}}
\end{table}

As shown in Table~\ref{tab:query_user}, refined queries receive consistently higher ratings. Participants judged them to be more specific and clinically relevant, supporting the claim that the MHVRC generation pipeline captures realistic multi-turn information needs.

The second study evaluates retrieval quality by comparing ground-truth videos with DATR Top-1 results. Participants rated procedural accuracy, instructional clarity, clinical suitability, and overall preference. Table~\ref{tab:retrieval_user} shows that DATR's Top-1 results are close to ground truth in all dimensions, indicating that the model can surface relevant instructional material even when exact ground-truth matching is difficult.

\begin{table}[t]
  \centering
  \caption{User evaluation of retrieval outputs.}
  \label{tab:retrieval_user}
  \resizebox{\linewidth}{!}{
  \begin{tabular}{lrrr}
    \toprule
    Dimension & Ground truth & DATR Top-1 & Difference\\
    \midrule
    Procedural accuracy & 4.7 & 4.2 & -0.5\\
    Instructional clarity & 4.6 & 4.3 & -0.3\\
    Clinical suitability & 4.5 & 4.1 & -0.4\\
    Overall preference & 4.6 & 4.2 & -0.4\\
    \bottomrule
  \end{tabular}}
\end{table}

\subsection{Ablation Study}
\begin{table*}[t]
  \centering
  \caption{Ablation results. Each block isolates one design choice in DATR.}
  \label{tab:ablation}
  \begin{tabular}{llrrrrr}
    \toprule
    Study & Setting & R@1 & R@5 & R@10 & MedR & MeanR\\
    \midrule
    Backbone & LSTM & 11.2 & 30.4 & 43.1 & 14 & 40.3\\
             & GRU & 12.0 & 31.9 & 44.7 & 13 & 38.9\\
             & Transformer & \textbf{19.5} & \textbf{44.8} & \textbf{57.1} & \textbf{7} & \textbf{25.6}\\
    \midrule
    Stage II & Without Stage II & 14.0 & 36.5 & 49.8 & 11 & 32.9\\
             & With Stage II & \textbf{19.5} & \textbf{44.8} & \textbf{57.1} & \textbf{7} & \textbf{25.6}\\
    \midrule
    Fusion & Add only & 17.1 & 40.3 & 52.9 & 9 & 29.5\\
           & Multiply only & 16.5 & 39.8 & 52.1 & 9 & 30.1\\
           & Add + Mul + MLP & \textbf{19.5} & \textbf{44.8} & \textbf{57.1} & \textbf{7} & \textbf{25.6}\\
    \midrule
    CLIP loss & Text-to-video only & 18.0 & 42.3 & 54.2 & 9 & 28.7\\
              & Bidirectional & \textbf{19.5} & \textbf{44.8} & \textbf{57.1} & \textbf{7} & \textbf{25.6}\\
    \bottomrule
  \end{tabular}
\end{table*}

Table~\ref{tab:ablation} summarizes the main ablations. Transformer encoding substantially outperforms recurrent backbones, confirming that self-attention is useful for long procedural videos. Removing Stage II causes a large drop, which indicates that coarse retrieval alone cannot capture refined query intent. For query fusion, additive-only fusion preserves broad topic continuity but underemphasizes new constraints; multiplicative-only fusion highlights differences but may lose the original context. Combining addition, multiplication, and an MLP gives the strongest performance. Finally, bidirectional contrastive learning improves alignment relative to text-to-video loss alone.

We also evaluate the scope of re-ranking. Full-database re-ranking obtains R@1 of 19.7 and R@10 of 57.1, while re-ranking only Stage-I top-100 candidates obtains R@1 of 19.5 and R@10 of 56.8 with 0.4$\times$ the computation. This small accuracy trade-off motivates the top-100 design in DATR.

\subsection{Error Analysis}
We observe three recurring error types. The first is topic-level confusion. These failures occur when the model retrieves a video from the same broad category but with the wrong procedural focus, such as returning a general lower-body exercise for a query about injury-prevention squat form. Stage II reduces but does not eliminate this issue because some visual differences are subtle and may require pose-level features.

The second error type is temporal-phase mismatch. Health videos often contain multiple steps, and a refined query may refer to a particular phase rather than the whole clip. A video can therefore be relevant in topic but weak in temporal localization. Our current benchmark treats retrieval at the video level, so the model may receive credit or penalty based on an entire clip even when only a segment is ideal. Future extensions could add moment-level annotations to evaluate whether the system retrieves the correct temporal span.

The third error type is over-specific refinement. Generated follow-up queries sometimes add a constraint that is visible but not central to the medical procedure, such as clothing, room layout, or background equipment. Multiplicative fusion can overemphasize these details if they appear rare in the candidate set. This explains why additive-only and multiplicative-only fusion are both weaker than the combined strategy: the model must preserve the core topic while using refinements selectively.

\subsection{Practical Use Cases}
Interactive retrieval is useful in several health-facing scenarios. In clinical training, instructors may search for demonstrations that match a particular patient posture, examination technique, or rehabilitation phase. In patient education, users may begin with a simple query and then refine it by pain location, required equipment, or desired difficulty. In remote rehabilitation, therapists may retrieve videos to support home exercises while adjusting constraints based on patient feedback. In medical misinformation mitigation, a dialogue-aware system could help users move from vague online searches toward more precise, trustworthy instructional material.

These use cases share a common requirement: the system must be responsive to evolving intent. A static ranking is not enough when the user learns what to ask only after interacting with the system. DATR is therefore best understood not as a replacement for expert judgment, but as an infrastructure component for more adaptive search interfaces.

\section{Discussion and Limitations}
DATR shows that multi-turn interaction can improve health video retrieval, but several limitations remain. First, MHVRC relies on generated descriptions and generated refinements. Although prompts are grounded in video content and user studies support their usefulness, expert verification is still needed before deployment in clinical settings. Second, the current setting focuses on two turns. Real search sessions may involve longer dialogues, corrections, and explicit feedback such as ``not this kind of exercise'' or ``show an easier version.'' Third, videos are represented through visual frames and generated text, while many health videos include audio narration, subtitles, on-screen labels, and motion trajectories that could provide additional evidence. Fourth, retrieval does not equal medical advice. A system may surface relevant instructional videos, but it should not determine diagnosis, contraindications, or treatment plans without professional oversight.

These limitations suggest future directions. Longer multi-turn datasets could model session-level rehabilitation or surgical training. Audio, subtitle, and pose features could improve multimodal grounding. Clinician-in-the-loop evaluation could measure safety, not only relevance. Finally, personalized retrieval could account for user capability and cognitive load, connecting health video search with emerging work on personalized human motion and dance synthesis \citep{personaldance2024}.

\section{Ethical and Safety Considerations}
Health video retrieval systems should be evaluated not only for relevance but also for safety. A retrieved video may be topically correct yet inappropriate for a user's condition, age, injury history, or clinical status. MHVRC does not authorize treatment recommendations; it supports retrieval research over instructional videos. Any deployed system should clearly communicate that retrieved content is informational and should be interpreted with professional guidance when medical decisions are involved.

The use of generated annotations also requires caution. Vision-language models can omit important details or describe visual content with unwarranted certainty. Language models can produce fluent refinements that sound clinically plausible even when the underlying evidence is weak. Our pipeline mitigates this risk by grounding refinements in descriptions, but human review remains important for high-stakes applications. In future versions of MHVRC, clinician validation could be used to label whether a refined query is safe, answerable, and supported by the video.

Privacy is another consideration. Although the videos in MHVRC are instructional and public-facing, health-related content can still involve sensitive demonstrations. Dataset builders should avoid content that reveals private patient information or stigmatizing conditions without consent. Retrieval systems should also avoid amplifying low-credibility or misleading videos. Integrating source credibility, clinical review, and uncertainty estimation would make interactive health video retrieval more responsible and practically useful.

\section{Conclusion}
We introduced interactive multi-turn health video retrieval, a task that better reflects how users search for clinically useful instructional content. We constructed MHVRC by pairing health videos with video-grounded descriptions and refined follow-up queries, and proposed DATR, a dialogue-aware two-stage retrieval framework. Experiments show that DATR outperforms strong single-turn baselines, while user studies confirm that refined queries are more specific, clinically relevant, and useful. The work provides a foundation for health video systems that retrieve not only what a user initially asks for, but what the user progressively comes to mean.

{\small
\bibliographystyle{ieee_fullname}
\bibliography{references}

@inproceedings{luo2021clip4clip,
  title={{CLIP4Clip}: An Empirical Study of {CLIP} for End to End Video Clip Retrieval and Captioning},
  author={Luo, Huaishao and Ji, Lei and Zhong, Ming and Chen, Yang and Lei, Wen and Duan, Nan and Li, Tianrui},
  booktitle={Advances in Neural Information Processing Systems},
  year={2021}
}

@inproceedings{bain2021frozen,
  title={Frozen in Time: A Joint Video and Image Encoder for End-to-End Retrieval},
  author={Bain, Max and Nagrani, Arsha and Varol, G{\"u}l and Zisserman, Andrew},
  booktitle={Proceedings of the IEEE/CVF International Conference on Computer Vision},
  pages={1728--1738},
  year={2021}
}

@inproceedings{radford2021clip,
  title={Learning Transferable Visual Models From Natural Language Supervision},
  author={Radford, Alec and Kim, Jong Wook and Hallacy, Chris and Ramesh, Aditya and Goh, Gabriel and Agarwal, Sandhini and Sastry, Girish and Askell, Amanda and Mishkin, Pamela and Clark, Jack and Krueger, Gretchen and Sutskever, Ilya},
  booktitle={Proceedings of the International Conference on Machine Learning},
  pages={8748--8763},
  year={2021}
}

@inproceedings{lei2021clipbert,
  title={Less is More: {CLIPBERT} for Video-and-Language Learning via Sparse Sampling},
  author={Lei, Jie and Lyu, Chenliang and Chen, Liangchen and Li, Yao and Lu, Xiaowu and Berg, Tamara L.},
  booktitle={Proceedings of the IEEE/CVF Conference on Computer Vision and Pattern Recognition},
  pages={7331--7341},
  year={2021}
}

@inproceedings{abacha2022medvidqa,
  title={{MedVidQA}: A Large-Scale Medical Video Question Answering Dataset},
  author={Abacha, Asma Ben and Yim, Wen-wai and Fan, Yujuan and Lin, Thomas and Demner-Fushman, Dina},
  booktitle={Proceedings of the Thirteenth Language Resources and Evaluation Conference},
  year={2022}
}

@inproceedings{xu2016msrvtt,
  title={{MSR-VTT}: A Large Video Description Dataset for Bridging Video and Language},
  author={Xu, Jun and Mei, Tao and Yao, Ting and Rui, Yong},
  booktitle={Proceedings of the IEEE/CVF Conference on Computer Vision and Pattern Recognition},
  pages={5288--5296},
  year={2016}
}

@inproceedings{zhou2018youcook2,
  title={Towards Automatic Learning of Procedures from Web Instructional Videos},
  author={Zhou, Luowei and Xu, Chenliang and Corso, Jason J.},
  booktitle={Proceedings of the AAAI Conference on Artificial Intelligence},
  pages={7590--7598},
  year={2018}
}

@inproceedings{lei2020tvr,
  title={{TVR}: A Large-Scale Dataset for Video-Subtitle Moment Retrieval},
  author={Lei, Jie and Yu, Licheng and Berg, Tamara L. and Bansal, Mohit},
  booktitle={Proceedings of the European Conference on Computer Vision},
  pages={447--463},
  year={2020}
}

@inproceedings{yu2020hero,
  title={{HERO}: Hierarchical Encoder for Video+Language Omni-representation Pre-training},
  author={Li, Linjie and Chen, Yen-Chun and Cheng, Yu and Gan, Zhe and Yu, Licheng and Liu, Jingjing},
  booktitle={Proceedings of the Conference on Empirical Methods in Natural Language Processing},
  pages={2046--2065},
  year={2020}
}

@inproceedings{krishna2017activitynet,
  title={Dense-Captioning Events in Videos},
  author={Krishna, Ranjay and Hata, Kenji and Ren, Frederic and Fei-Fei, Li and Niebles, Juan Carlos},
  booktitle={Proceedings of the IEEE/CVF International Conference on Computer Vision},
  pages={706--715},
  year={2017}
}

@inproceedings{hendricks2017didemo,
  title={Localizing Moments in Video with Natural Language},
  author={Hendricks, Lisa Anne and Wang, Oliver and Shechtman, Eli and Sivic, Josef and Darrell, Trevor and Russell, Bryan},
  booktitle={Proceedings of the IEEE/CVF International Conference on Computer Vision},
  pages={5804--5813},
  year={2017}
}

@inproceedings{lei2018tvqa,
  title={{TVQA}: Localized, Compositional Video Question Answering},
  author={Lei, Jie and Yu, Licheng and Berg, Tamara L. and Bansal, Mohit},
  booktitle={Proceedings of the Conference on Empirical Methods in Natural Language Processing},
  pages={1369--1379},
  year={2018}
}

@article{yu2021cir,
  title={A Survey of Conversational Search},
  author={Gao, Jianfeng and Xue, Chenyan and Dong, Anlei and Chen, Jiafeng},
  journal={Foundations and Trends in Information Retrieval},
  volume={14},
  number={5},
  pages={371--490},
  year={2021}
}

@article{nogueira2019bert,
  title={Passage Re-ranking with {BERT}},
  author={Nogueira, Rodrigo and Cho, Kyunghyun},
  journal={arXiv preprint arXiv:1901.04085},
  year={2019}
}

@inproceedings{bertasius2021timesformer,
  title={Is Space-Time Attention All You Need for Video Understanding?},
  author={Bertasius, Gedas and Wang, Heng and Torresani, Lorenzo},
  booktitle={Proceedings of the International Conference on Machine Learning},
  year={2021}
}

@inproceedings{arnab2021vivit,
  title={{ViViT}: A Video Vision Transformer},
  author={Arnab, Anurag and Dehghani, Mostafa and Heigold, Georg and Sun, Chen and Lu\v{c}i\'c, Mario and Schmid, Cordelia},
  booktitle={Proceedings of the IEEE/CVF International Conference on Computer Vision},
  pages={6836--6846},
  year={2021}
}

@article{twinanda2017endonet,
  title={{EndoNet}: A Deep Architecture for Recognition Tasks on Laparoscopic Videos},
  author={Twinanda, Andru P. and Shehata, Sherif and Mutter, Didier and Marescaux, Jacques and de Mathelin, Michel and Padoy, Nicolas},
  journal={IEEE Transactions on Medical Imaging},
  volume={36},
  number={1},
  pages={86--97},
  year={2017}
}

@article{jin2018svrcnet,
  title={{SV-RCNet}: Workflow Recognition From Surgical Videos Using Recurrent Convolutional Network},
  author={Jin, Yueming and Dou, Qi and Chen, Hao and Yu, Lequan and Qin, Jing and Fu, Chi-Wing and Heng, Pheng-Ann},
  journal={IEEE Transactions on Medical Imaging},
  volume={37},
  number={5},
  pages={1114--1126},
  year={2018}
}

@article{bae2019rehab,
  title={Automatic Exercise Assessment in Physical Rehabilitation},
  author={Bae, Sujin and Kim, Jooyeon and Park, Jihye and Lee, Sangyoun},
  journal={Sensors},
  volume={19},
  number={19},
  pages={4113},
  year={2019}
}

@article{li2020smartrehab,
  title={Smart Rehabilitation Based on Artificial Intelligence and Internet of Things: A Survey},
  author={Li, Yong and Hu, Jie and Zhang, Yu},
  journal={IEEE Access},
  volume={8},
  pages={180246--180271},
  year={2020}
}

@article{demner2009clinical,
  title={What Can Natural Language Processing Do for Clinical Decision Support?},
  author={Demner-Fushman, Dina and Chapman, Wendy W. and McDonald, Clement J.},
  journal={Journal of Biomedical Informatics},
  volume={42},
  number={5},
  pages={760--772},
  year={2009}
}

@article{wang2023videorehab,
  title={Computer Vision for Musculoskeletal Rehabilitation: A Survey},
  author={Wang, Jiang and Liu, Xinyi and Jiang, Zhenyu and Zhang, Qi},
  journal={IEEE Reviews in Biomedical Engineering},
  year={2023}
}

@inproceedings{tang2023motionbert,
  title={{MotionBERT}: A Unified Perspective on Learning Human Motion Representations},
  author={Zhu, Wentao and Ma, Xiaoxuan and Liu, Zhaoyang and Liu, Libin and Wu, Wayne and Wang, Yizhou},
  booktitle={Proceedings of the IEEE/CVF International Conference on Computer Vision},
  year={2023}
}

@article{chang2023magic,
  title={{MagicPose}: Realistic Human Poses and Facial Expressions Retargeting with Identity-aware Diffusion},
  author={Chang, Di and Shi, Yichun and Gao, Quankai and Xu, Jiawei and Fu, Hongbo},
  journal={arXiv preprint arXiv:2311.12052},
  year={2023}
}

@inproceedings{hu2023animate,
  title={Animate Anyone: Consistent and Controllable Image-to-Video Synthesis for Character Animation},
  author={Hu, Li and Gao, Xin and Zhang, Peng and Sun, Ke and Zhang, Bang and Bo, Liefeng},
  booktitle={Proceedings of the IEEE/CVF Conference on Computer Vision and Pattern Recognition},
  year={2024}
}

@article{yang2024megadance,
  title={{MEGADance}: Mixture-of-Experts Architecture for Genre-Aware 3D Dance Generation},
  author={Yang, Kaixing and Tang, Xulong and Peng, Ziqiao and Hu, Yuxuan and He, Jun and Liu, Hongyan},
  journal={arXiv preprint arXiv:2505.17543},
  year={2025}
}

@article{yang2024flowerdance,
  title={{FlowerDance}: {MeanFlow} for Efficient and Refined 3D Dance Generation},
  author={Yang, Kaixing and Tang, Xulong and Peng, Ziqiao and Zhang, Xiangyue and Wang, Puwei and He, Jun and Liu, Hongyan},
  journal={arXiv preprint arXiv:2511.21029},
  year={2025}
}

@article{zhang2024bitdiff,
  title={{BiTDiff}: Fine-Grained 3D Conducting Motion Generation via {BiMamba}-Transformer Diffusion},
  author={Jia, Tianzhi and Yang, Kaixing and Yang, Xiaole and Tang, Xulong and Qiu, Ke and Wei, Shikui and Zhao, Yao},
  journal={arXiv preprint arXiv:2604.04395},
  year={2026}
}

@inproceedings{zhang2024dancecamera3d,
  title={{DanceCamera3D}: 3D Camera Movement Synthesis with Music and Dance},
  author={Wang, Zixuan and Jia, Jia and Sun, Shikun and Wu, Haozhe and Han, Rong and Li, Zhenyu and Tang, Di and Zhou, Jiaqing and Luo, Jiebo},
  booktitle={Proceedings of the IEEE/CVF Conference on Computer Vision and Pattern Recognition},
  pages={7892--7901},
  year={2024}
}

@inproceedings{zhang2023tm2d,
  title={{TM2D}: Bimodality Driven 3D Dance Generation via Music-Text Integration},
  author={Gong, Kehong and Lian, Defu and Chang, Heng and Guo, Chuan and Jiang, Zihang and Zuo, Xinxin and Mi, Michael Bi and Wang, Xinchao},
  booktitle={Proceedings of the IEEE/CVF International Conference on Computer Vision},
  year={2023}
}

@inproceedings{yang2024beatdance,
  title={{BeatDance}: A Beat-Based Model-Agnostic Contrastive Learning Framework for Music-Dance Retrieval},
  author={Yang, Kaixing and Zhou, Xukun and Tang, Xulong and Diao, Ran and Liu, Hongyan and He, Jun and Fan, Zhaoxin},
  booktitle={Proceedings of the 2024 International Conference on Multimedia Retrieval},
  pages={11--19},
  year={2024}
}

@inproceedings{jiang2023motiongpt,
  title={{MotionGPT}: Human Motion as a Foreign Language},
  author={Jiang, Biao and Chen, Xin and Liu, Wen and Yu, Jingyi and Yu, Gang and Chen, Tao},
  booktitle={Advances in Neural Information Processing Systems},
  year={2023}
}

@inproceedings{guo2022tm2t,
  title={Generating Diverse and Natural 3D Human Motions from Text},
  author={Guo, Chuan and Zuo, Xinxin and Wang, Sen and Zou, Shihao and Sun, Qingyao and Deng, Annan and Gong, Minglun and Cheng, Li},
  booktitle={Proceedings of the IEEE/CVF Conference on Computer Vision and Pattern Recognition},
  pages={5152--5161},
  year={2022}
}

@inproceedings{li2021ai,
  title={{AI Choreographer}: Music Conditioned 3D Dance Generation with {AIST++}},
  author={Li, Ruilong and Yang, Shan and Ross, David A. and Kanazawa, Angjoo},
  booktitle={Proceedings of the IEEE/CVF International Conference on Computer Vision},
  pages={13401--13412},
  year={2021}
}

@inproceedings{siyao2022bailando,
  title={{Bailando}: 3D Dance Generation by Actor-Critic {GPT} with Choreographic Memory},
  author={Siyao, Li and Yu, Weijiang and Gu, Tianpei and Lin, Chunze and Wang, Quan and Qian, Chen and Loy, Chen Change and Liu, Ziwei},
  booktitle={Proceedings of the IEEE/CVF Conference on Computer Vision and Pattern Recognition},
  year={2022}
}

@article{yang2025macedance,
  title={{MACE-Dance}: Motion-Appearance Cascaded Experts for Music-Driven Dance Video Generation},
  author={Yang, Kaixing and Zhu, Jiashu and Tang, Xulong and Peng, Ziqiao and Zhang, Xiangyue and Wang, Puwei and Wu, Jiahong and Chu, Xiangxiang and Liu, Hongyan and He, Jun},
  journal={arXiv preprint arXiv:2512.18181},
  year={2025}
}

@article{uniavgen2025,
  title={{UniAVGen}: Unified Audio and Video Generation with Asymmetric Cross-Modal Interactions},
  author={Zhang, Guozhen and Zhou, Zixiang and Hu, Teng and Peng, Ziqiao and Zhang, Youliang and Chen, Yi and Zhou, Yuan and Lu, Qinglin and Wang, Limin},
  journal={arXiv preprint arXiv:2511.03334},
  year={2025}
}

@inproceedings{zhang2025semtalk,
  title={{SemTalk}: Holistic Co-speech Motion Generation with Frame-level Semantic Emphasis},
  author={Zhang, Xiangyue and Li, Jianfang and Zhang, Jiaxu and Dang, Ziqiang and Ren, Jianqiang and Bo, Liefeng and Tu, Zhigang},
  booktitle={Proceedings of the IEEE/CVF International Conference on Computer Vision},
  pages={13761--13771},
  year={2025}
}

@inproceedings{zhang2025echomask,
  title={{EchoMask}: Speech-Queried Attention-based Mask Modeling for Holistic Co-Speech Motion Generation},
  author={Zhang, Xiangyue and Li, Jianfang and Zhang, Jiaxu and Ren, Jianqiang and Bo, Liefeng and Tu, Zhigang},
  booktitle={Proceedings of the ACM International Conference on Multimedia},
  pages={10827--10836},
  year={2025}
}

@article{zhang2025robust2d,
  title={Robust {2D} Skeleton Action Recognition via Decoupling and Distilling {3D} Latent Features},
  author={Zhang, Xiangyue and Pan, Kunkun and Wang, Di and Jiang, Xinchen and Tu, Zhigang},
  journal={IEEE Transactions on Circuits and Systems for Video Technology},
  year={2025}
}

@inproceedings{tevet2023mdm,
  title={Human Motion Diffusion Model},
  author={Tevet, Guy and Raab, Sigal and Gordon, Brian and Shafir, Yonatan and Cohen-Or, Daniel and Bermano, Amit H.},
  booktitle={International Conference on Learning Representations},
  year={2023}
}

@inproceedings{chen2023executing,
  title={Executing your Commands via Motion Diffusion in Latent Space},
  author={Chen, Xin and Jiang, Biao and Liu, Wen and Huang, Zilong and Fu, Bin and Chen, Tao and Yu, Gang},
  booktitle={Proceedings of the IEEE/CVF Conference on Computer Vision and Pattern Recognition},
  year={2023}
}

@inproceedings{personaldance2024,
  title={Personalized Dance Synthesis Based on Physical and Cognitive Intensities},
  author={Tang, Xulong and Yeo, Eun and Mao, Ruiyu and Guo, Xiaohu and Alghofaili, Rawan},
  booktitle={Proceedings of the IEEE Conference on Virtual Reality and 3D User Interfaces},
  pages={261--271},
  year={2026}
}
}

\end{document}